# Title: Repeated deterministic defect-assisted switching of magnetic vortex core polarity

**Authors:** Mahdi Mehrnia[1], Jeremy Trimble[1], Olle Heinonen[2], Jesse Berezovsky[1*]

**Affiliations:**

[1]Department of Physics, Case Western Reserve University, 10900 Euclid Avenue, Cleveland, OH 44106, USA.

[2]Materials Science Division, Argonne National Laboratory, Lemont, Illinois 60439, USA.

Corresponding author: E-mail: jab298@case.edu

**Abstract:** Because of its stability, the polarity of a magnetic vortex core (VC) is a candidate for binary data storage. Switching can be accomplished, e.g, by driving the VC above a critical velocity $v_c$. Here, we report on controlled and repeated switching of VC polarity by significantly reducing $v_c$ locally. We excite vortex dynamics in thin Permalloy disks with a magnetic field pulse, and map the two-dimensional VC trajectory using time-resolved Kerr microscopy. In pristine samples, we observe normal gyrotropic motion of the VC. After laser-induced generation of defects, however, we observe repeated VC reversal at much-reduced critical velocities as low as 20 m/s. Micromagnetic simulations reveal how local reduction of exchange coupling can create VC reversal sites for deterministic VC switching.

**One Sentence Summary:** Magnetic vortex polarity, normally stable due to topology, is readily reversed when the core traverses nanoscale defect regions.

The large energy barrier between the two polarity states of a magnetic vortex core (VC) makes it stable against a polarity switch, such that an out-of-plane static field of about 0.5 T is required for this purpose (*1*). The stability of topologically-protected vortices and related magnetic textures such as skyrmions makes them potential candidates for data storage devices (*2*). For applications, we require a method to switch the polarity that is fast, efficient, and deterministic. In previous work, VC polarity has been switched by either applying a large magnetic field or by driving the VC to a critical velocity $v_c \cong 1.66\,\gamma\sqrt{A}$ ($\gamma$ is the gyromagnetic ratio, and $A$ is the exchange stiffness), which is $\sim 350\ m/s$ for Permalloy (*3*, *4*). High-speed switching by exceeding $v_c$ has been demonstrated by various methods (*5–10*).

In this study, we experimentally observe deterministic VC switching at $v_c$ as low as 20 $m/s$ when the VC passes through nanoscale localized regions of the material. We attribute this switching to a local reduction of $A$ at laser-induced defects. By tuning the excitation pulse and bias magnetic field, we can control the number of switches per pulse from zero up to eight, as well as demonstrate deterministic single-shot switching.

The characteristic magnetic texture in micron-sized soft ferromagnetic disks is a vortex state, in which the magnetization curls around in-plane, see Fig. 1(**A**). However, due to short range exchange interaction, magnetization starts to turn out of the plane at the center of the vortex in a region $\sim 10$ nm. This region, the vortex core, constitutes a topological defect with polarity $p = \pm 1$, either up or down. An in-plane magnetic field sets the VC equilibrium position $x_0 \approx$

$\chi_0(B_y, -B_x)$, with $\chi_0 \approx 70$ nm/mT here. By applying a fast magnetic field step, the VC undergoes dynamics out of equilibrium, following a gyrating trajectory $x(t)$ governed by the Thiele equation (*11*), eventually spiraling in to the new $x_0$ on a tens of nanosecond timescale. The sense of VC gyration, clockwise (CW) or counter-clockwise (CCW), is determined by the polarity $p = \pm 1$ of the VC. Throughout this paper, we consider CW and CCW gyrations as "core down" (shown by solid lines) and "core up" (shown by dashed lines) respectively.

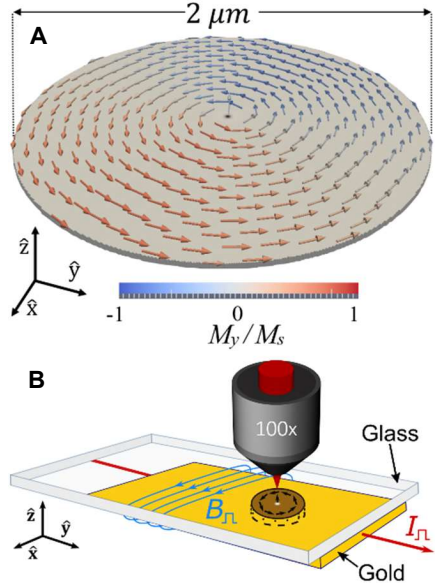

**Fig. 1. Samples and the magnetic vortex.** (**A**) A simulated ferromagnetic vortex state with the core polarity up. (**B**) Schematic of the sample, a Permalloy disk (brown) is capped by a gold coplanar waveguide. A microscope objective is used to focus the laser light on the disks

We fabricate 2.2-$\mu m$ diameter, 40-$nm$ thick permalloy disks with an adjacent coplanar waveguide to excite dynamics of a disk's vortex state (Fig. 1(**B**)). These dynamics are detected using a 3D time-resolved magneto-optical Kerr effect (3D TR-MOKE) technique (*12*) (see Supplementary Material.)

As we will see below, defects in a ferromagnetic material play an important role in determining VC dynamics. In order to characterize the positions and strengths of these defects, we use the VC as a probe, raster scanning an in-plane magnetic field $B_0$, and observing the effect of defects on the vortex equilibrium position $x_0$. When the vortex core becomes pinned at a defect, we observe a plateau in $x_0$ vs. $B_0$, and often hysteretic $x_0$ depending on the sweep direction of $B_0$. The length of the plateaus, the size of jumps in and out of the plateaus, and the degree of hysteresis reveal the spatial extent and strength of the pinning (*13*). We will use this method to observe the enhancement of pinning due to laser exposure, and correlate it with the observation of VC switching.

We now compare the VC gyrations before and after prolonged exposure of the pulsed laser. We demonstrate that after extended laser exposure, VC reversal occurs at multiple spots. Figures 2(**A**) and (**B**) show the VC gyrations in response to a 35 ns magnetic pulse before and after prolonged laser exposure time. In Fig. 2(**A**), the scan begins before the pulse ($\Delta t < 0$) when $x(t) = x_0$. During the pulse, we observe gyrations about $x_1$. By the end of the pulse, the VC has partially relaxed to $x_1$. After the pulse, the VC again undergoes gyrations as it relaxes back to $x_0$. The sense of the gyrations is CCW everywhere (indicated by dashed lines), indicating a constant VC polarity with no polarity switch.

After further exposure to the pulsed laser (~ 6 hours) focused on the disk center, we observe significantly altered gyrations, Fig. 2(**B**). During the pulse, we observe CCW gyrations about $x_1$. Then as the VC approaches $(x, y) = (-32, -155)$ $nm$, the gyration switches to CW, indicating a core polarity switch. After the pulse ends, the VC begins relaxing to $x_0$, still with CW gyrations until it hits another switching spot near $(x, y) = (5, 90)$ $nm$ and suddenly reverses direction. As the VC continues to relax back to the center, we observe another seven switching spots, indicated by red arrows. Due to the stroboscopic nature of the experiment, each data point in the scan is an average of ~ $10^6$ repetitions of $x(t)$, indicating the deterministic nature of the trajectory, and the entire scan is taken multiple times to check the reproducibility of the measurement. The switching occurs mainly near the center of disk. Vortex dynamics away from the center show little to no switching, as discussed in the Supplementary Material.

Previous work has found that a critical VC velocity $v_c \approx 350 \ m/s$ is necessary to switch the polarity in permalloy samples. However, in Fig. 2(**B**), we observe VC reversal at velocities as low as $v_c = 20 \ m/s$, with switching observed for field pulses down to 1.6 mT amplitude (see Fig. S4 in supplementary material).

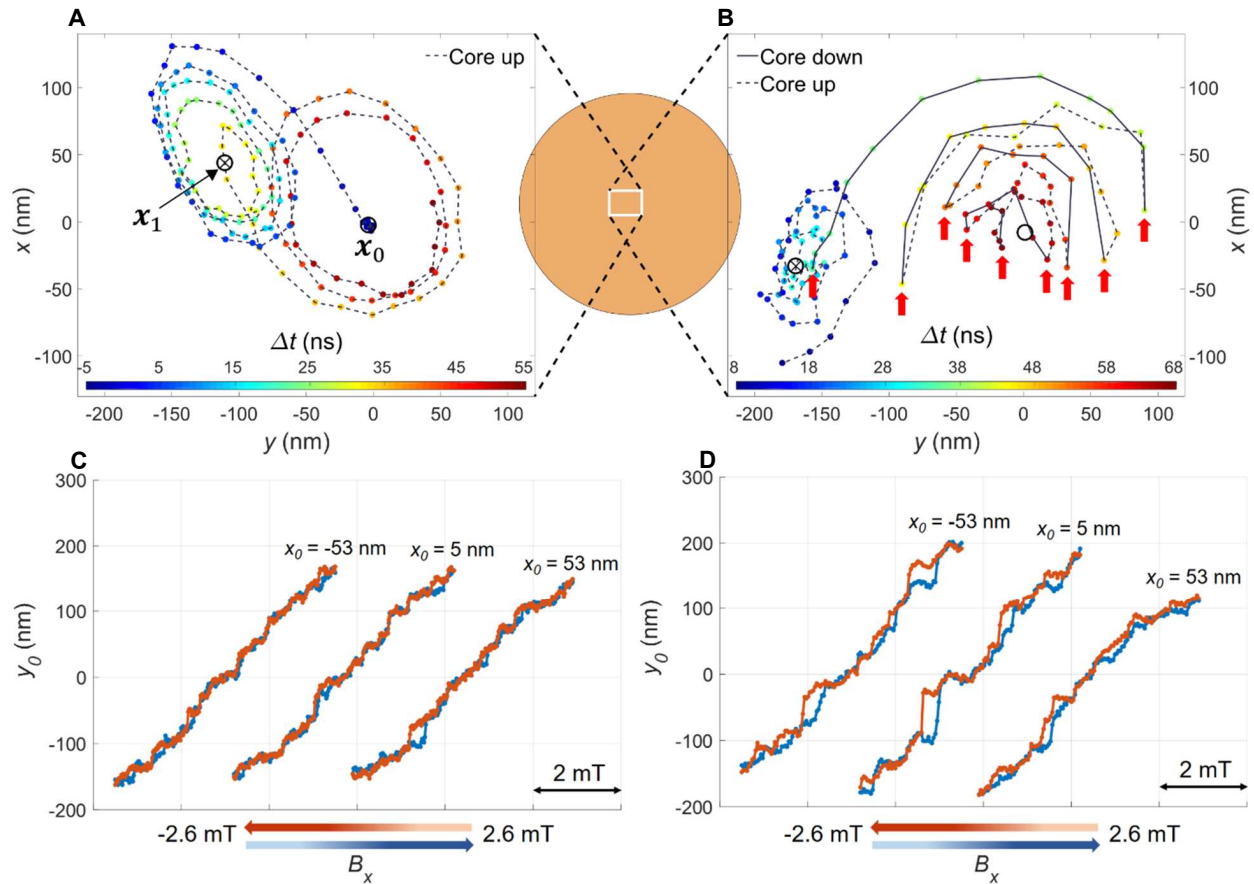

**Fig. 2. Magnetic vortex core reversal by defects. Each red arrow represents a VC reversal.** (**A**) and (**B**) Gyrations of the VC before and after prolonged pulsed laser exposure. The color bar $\Delta t$ indicates the time relative to the beginning of the pulse. (**C**) and (**D**) show the VC equilibrium position $y_0$ vs. $B_x$ swept up (blue) and down (red) at three different values of $x_0$ before and after the laser exposure respectively.

To understand how the laser has changed the defect landscape, we look at the $y_0$ vs. $B_x$ hysteresis loops at different values of $x_0$ (different $B_y$) before and after prolonged exposure. In the absence of any defects, we expect a linear $y_0 = \chi_0 B_x$. However, defects likely associated with grain boundaries or other structural inhomogeneities cause the plateaus, jumps, and hysteresis seen in Fig. 2(**C**) and (**D**) (*13, 14*). Figure 2(**C**) shows three hysteresis loops on the same sample before the prolonged exposure, offset horizontally for clarity. In these hysteresis loops, localized intrinsic defects created during the fabrication process cause small plateaus to appear with some hysteresis visible. Figure 2(**D**) shows the same measurement after ~ 13 hours of exposure to the pulsed laser. Clearly, we now observe more prominent plateaus, jumps, and hysteresis. We hypothesize that water or oxygen is trapped between the permalloy and the gold cap during the separate fabrication steps, which then undergoes laser-assisted diffusion into existing grain boundaries, resulting in local reductions in $A$ and/or $M_s$ (*15–18*).

For data storage applications, we require a single-shot switch with high reliability. Figure 3(**A**) illustrates a protocol for single-shot switching and subsequent polarity measurement. First, a single magnetic field pulse drives half a period of gyrotropic motion (green) towards a single switching site. Then the VC is adiabatically translated (purple) to another position where clear gyrations with no switching are observed (blue). The sense of these gyrations indicate the VC polarity. (See Supplementary Material for more detail.)

Figure 3(**B**), from top to bottom demonstrates the VC polarities after applying the single 5 $ns$ magnetic field pulses. Each core polarity up (down) is represented by a red up (blue down) arrow. Above each arrow, a green check-mark indicates a successful switch. Failed switches are represented by a red X. The polarity starts out up. Following one single-shot pulse, the polarity has flipped to down. This pattern repeats, and we verify that the polarity does not flip with no pulse, and flips twice with two pulses. From eighteen tests in this process, fifteen of them are successful. We believe the last two failures are due to further laser-induced enhancement of the defect after ~ one hour of measurement time.

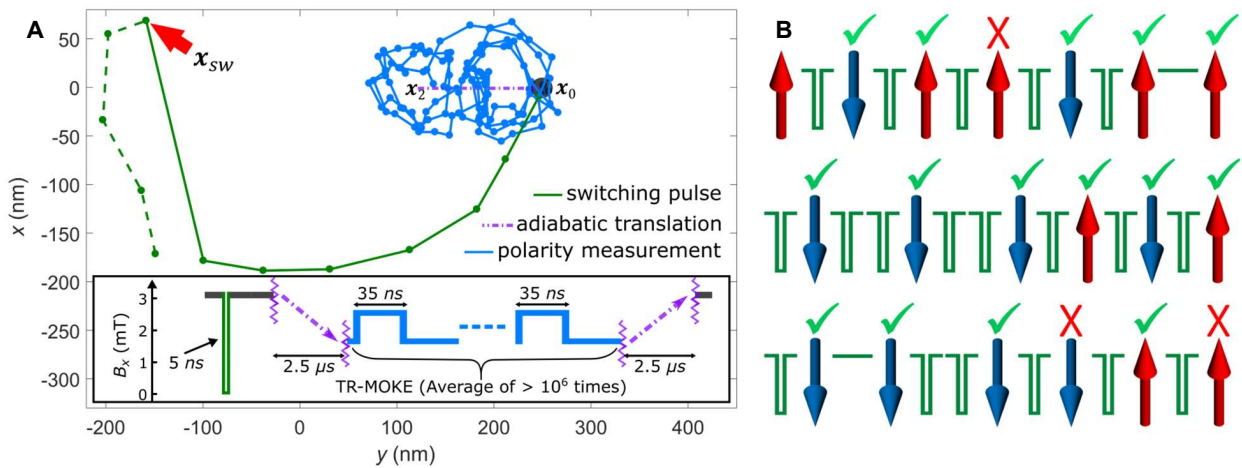

**Fig. 3 Single-shot switching.** (**A**) Schematic of the procedure used to switch and test the polarity. Red arrow indicates the reversal spot. (**B**) Core polarities before and after each switching pulse. Green check marks (red X) above each core polarity illustrate the successful (failed) switches.

To reveal the mechanism for the switching observed here, we simulate the interaction of the vortex core with a pinning site modelled as a localized region of reduced exchange constant $A$ and/or saturation magnetization $M_s$ (see Supplementary Material for details). Fig. 4(**A**) shows the simulated VC displacement $y_0$ vs. static field $B_x$ swept up and down, with $A$ reduced from 13 pJ/m to 10 pJ/m and 2.5 pJ/m in a circular region with a diameter of 30 nm. As $A$ is reduced, we observe the emergence of similar pinning and hysteresis as seen in the experiment following prolonged laser exposure (Fig. 2C and 2D).

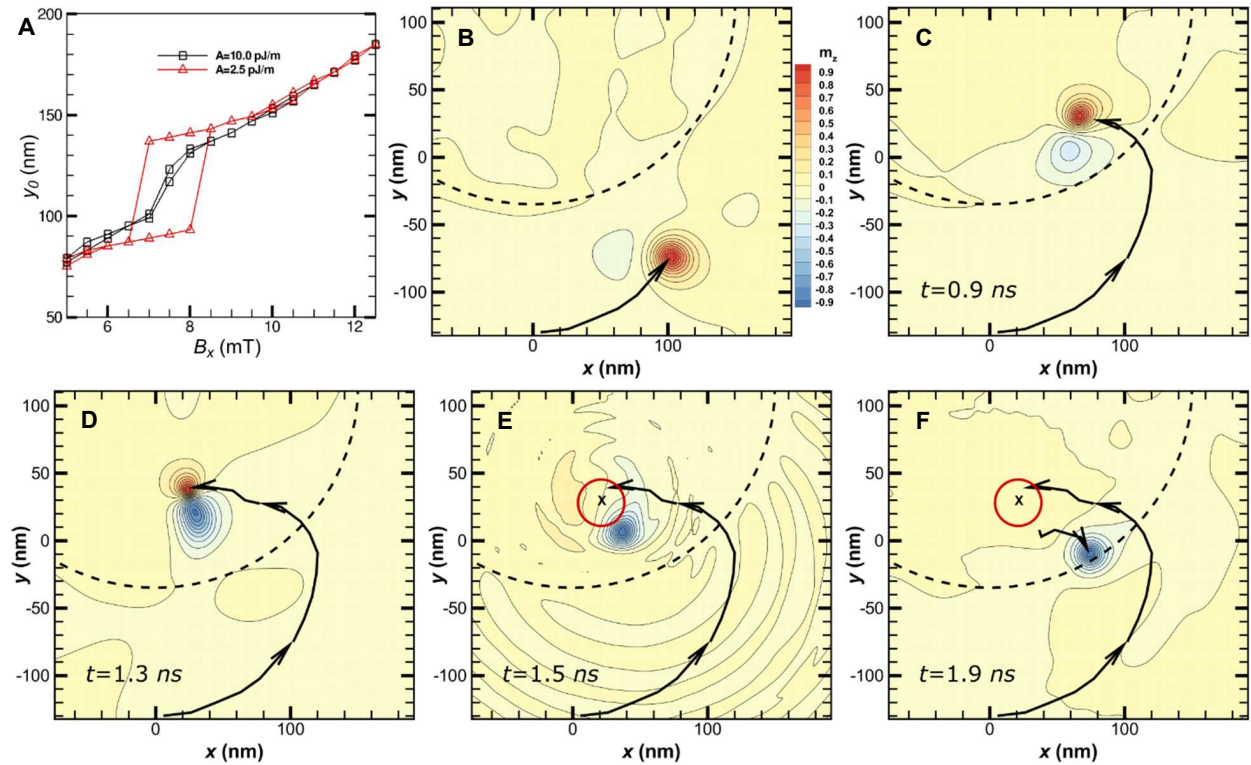

**Figure 4. Simulated defect-assisted VC reversal. (A)** VC equilibrium position $y_0$ vs. $B_x$ swept up and down. Pinning and hysteresis occur as the VC traverses a 30-nm diameter region where the exchange coupling is reduced to 10 pJ/m (black squares) and to 2.5 pJ/m (red triangles) from the nominal value of 13 pJ/m. **(B) – (F)** Snapshots of VC dynamics. **(B)** The VC (red) moves CCW towards a defect region shown as a dashed circle, with a slight blue wake visible behind the VC. **(C)** As the VC enters the defect region, the blue wake is significantly enhanced. **(D)** The wake has grown to be a new, reversed VC (blue). **(E)** The initial core is annihilated in a burst of spin wave emission. **(F)** The reversed core then continues on a CW trajectory.

We then simulate the non-equilibrium vortex dynamics with a trajectory passing through a defect region. For a reduction of $A$ by a factor of at least 4, we observe switching of the VC polarity as the VC passes through the low-$A$ region, and subsequent reversal of the VC precession. Figures 4(**B**) – (**F**) show a succession of time points as the vortex core approaches the defect region (dashed circle), as the VC is switching, and after the VC has switched polarity and precession direction (see Supplementary Material for additional details and videos).

The micromagnetic simulations support the hypothesis that defects serve as localized sites for rapid VC reversal with a significantly reduced critical velocity. The pattern of repeated reversals observed in the experiment suggest that these defects have an extended structure, likely due to grain boundaries, and possibly enhanced due to strain during fabrication or mounting. If VC polarity is to serve as a bit in future memory or logic devices, it will be important to control this defect-assisted switching process either to suppress bit-flip errors, or to engineer local sites for fast, reliable switching.

**Acknowledgements:** O.H. acknowledges funding from the US Department of Energy, Office of Science, Basic Energy Sciences Division of Materials Sciences and Engineering. We gratefully acknowledge the computing resources provided on Bebop and Blues, high-performance computing clusters operated by the Laboratory Computing Resource Center at Argonne National Laboratory.

**Supplementary Materials:**

Materials and Methods
Figures S1-S8
Movies S1-S2

## Supplementary material

**Materials and Methods:**

Samples in this study are 2.2-$\mu m$ diameter, 40-$nm$ thick permalloy disks, and are fabricated by electron beam lithography, electron beam evaporation, and liftoff, on a glass coverslip as illustrated in Fig. 1(**B**) in the main text. A 120-nm thick gold coplanar-waveguide (CPW) is fabricated on the glass coverslip over the disks. Electrical current pulses $I_\Pi$ through the gold CPW induce in-plane magnetic field pulses $\boldsymbol{B}_\Pi$ on the permalloy disks.

In order to observe the VC trajectory $\boldsymbol{x}(t)$, we use a 3D time-resolved magneto-optical Kerr effect (3D TR-MOKE) technique (*12*). In this method, a pulsed laser with 5 MHz repetition rate, $\lambda = 660\ nm$, pulse duration $\approx 20$ ps, and average power of 90 $\mu W$ before the 100x oil-immersion objective lens is focused on the disks after passing through the glass coverslip. The laser spot is centered on the VC equilibrium position and any motion of the VC within the probe spot results in a differential change in the net magnetization under the spot, and is detected as a change in the reflected polarization. In order to observe the gyrations of the VC, we excite the dynamics by 35-ns-duration magnetic field pulses with a variable delay $\Delta t$ relative to the laser pulses. We measure the equilibrium VC position $\boldsymbol{x_0}$ using static MOKE and raster scanning the sample past the probe laser (*12*).

**Additional VC reversal measurements:**

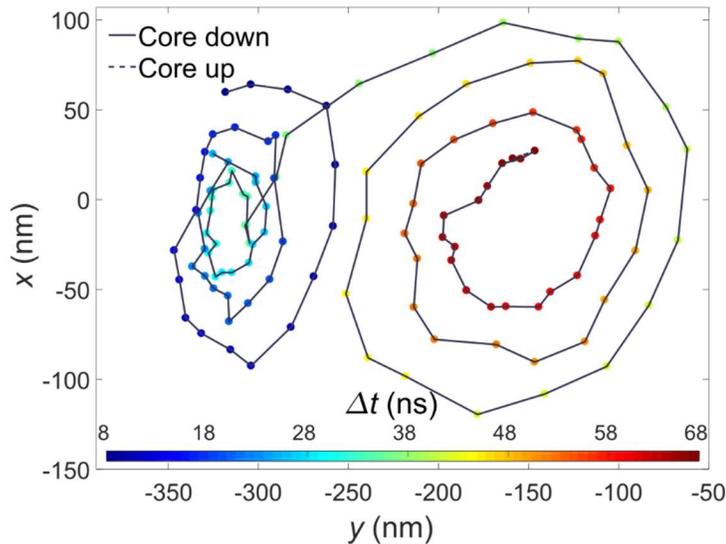

**Fig. S1.** The VC gyrations on left side of the disk after ~1 hour of pulsed laser exposure. The color bar $\Delta t$ shows the time relative to the beginning of the 35 $ns$ pulse.

Additional insight into the VC reversal mechanism is gained by translating the region of the VC dynamics away from the disk center using a bias field $\boldsymbol{B_0}$. Figure 2(**A**) in the main text shows VC dynamics near the center of the disk before prolonged laser exposure. In Fig. S1, we have translated the initial equilibrium position $x_0$ about 160 nm to the left with a bias field $B_0 = 2.1$ mT and measured the VC dynamics after about 1 hour of additional laser exposure. Here we see gyrations with little to no switching. (There may be a single reversal observed near the very end of the scan.)

Figure 2(**B**) in the main text shows VC dynamics centered on the disk after prolonged laser exposure at the disk center, where multiple switches are observed. In Figure S2, we show additional measurements at different bias fields $\boldsymbol{B_0}$ as the trajectory is centered away from to the center of the disk. The top panel shows the largest offset, and little to no switching is observed. This demonstrates that the switching behavior is mainly limited to the center of the disk, where laser exposure was greatest. The second panel shows the trajectory shifted slightly to the right, but still with little to no switching observed. The change in shape of the gyrations may be an artifact due to alignment of the probe laser. (If the VC moves toward the edge of the probe laser spot, the measured signal is less sensitive to VC displacement, resulting in a flattening of the measured displacement.) The third panel shows an intermediate regime where we believe the switching is not deterministic, discussed further below. Here we see a partial cancellation of the motion in the x-direction. Moving further towards the center, the fourth panel shows clear gyrations again, now with several switches visible. Finally, the bottom panel shows similar data as in Fig. 2(**B**) of the main text, with multiple VC reversals.

The right hand column of Fig. S2 shows additional data related to the z-component of magnetization $M_z$ provided by the 3D-MOKE measurement. The x- and y- components of the measurement represent the changes $\Delta M_x$ and $\Delta M_y$ of the net magnetization under the probe spot, and can be converted into spatial displacement of the VC. The $\Delta M_z$ component likewise represents the change in the out-of-plane component of magnetization convolved with the probe spot profile. Because the VC has a large out-of-plane component, the $\Delta M_z$ measurement is partially set by how the VC moves across the probe profile. However, longer-range out-of-plane distortions of the magnetization can also affect $\Delta M_z$ (even though the out-of-plane magnitude of these features is much less than that of the core, the convolution with the probe profile may still be significant because the core is much smaller than the probe spot.) The interpretation of the $\Delta M_z$ signal depends on the precise alignment of the probe spot with the VC dynamics. However, one generally expects to see oscillations as the VC gyrates, as seen in the top panels in Fig. S2. In the bottom panels, where VC reversal occurs, sharp kinks and jumps are seen in the $\Delta M_z$ data due to sudden reversals of the VC polarity and precession direction.

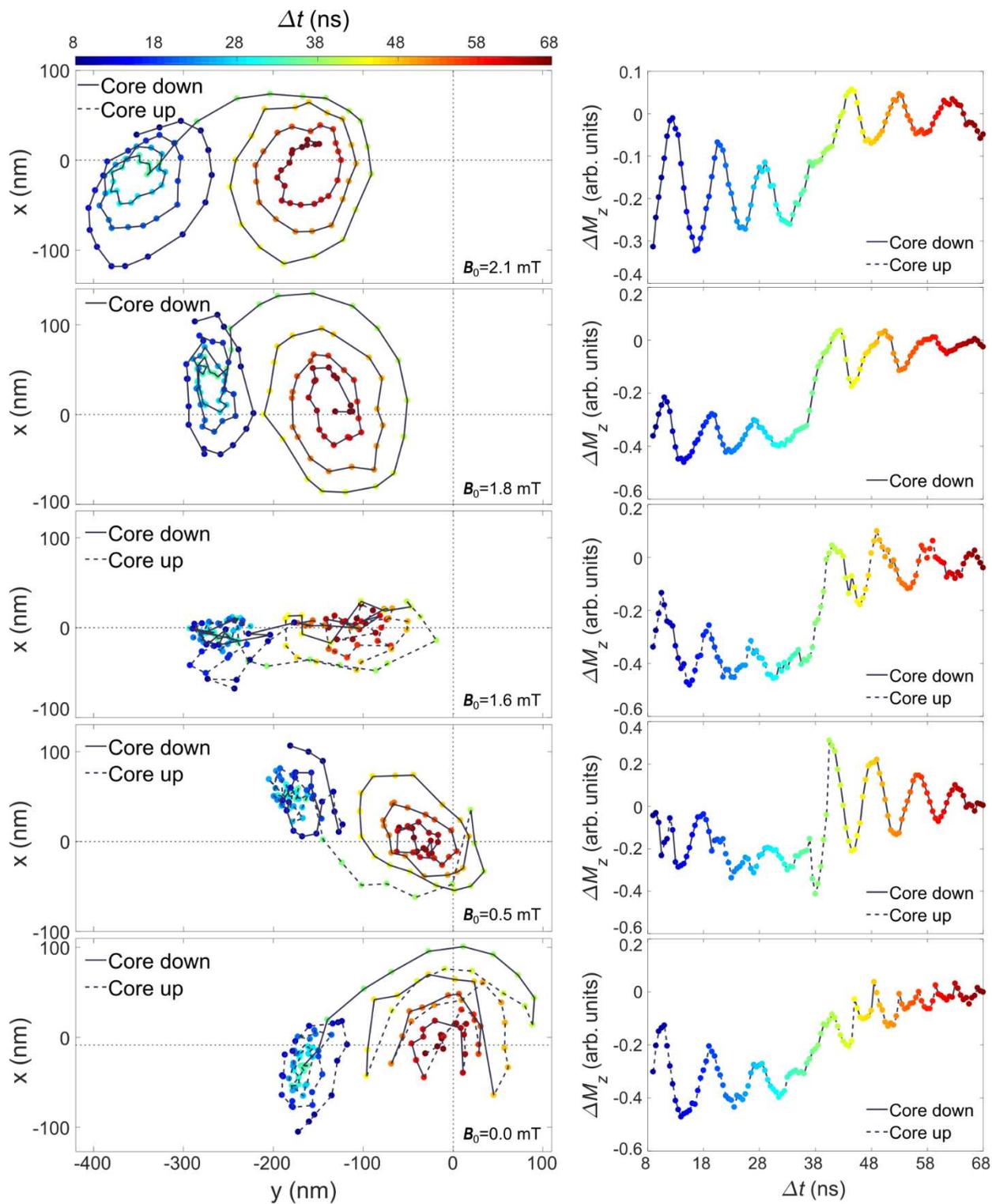

**Fig. S2.** From top to the bottom: 3D TR-MOKE of magnetization dynamics of the first sample as we shift the VC from the left region towards the center. Panels on the left demonstrate the 2D images for the VC gyrations and the panels on the right illustrate the change in the z-component. In all cases we used a 3.2 mT amplitude, 35 ns magnetic field pulse. We used fixed biased fields

($B_0$) to shift the VC from left (top panel) towards the center (bottom panel). Color bars $\Delta t$ indicate the time relative to the beginning of the pulse.

We further investigate the region of $B_0$ seen in the third panel of Fig. S2 (at $B_x = 1.6$ mT) attributed to stochastic switching. In Fig. S3, we show the trajectory at $B_x = 0.6$ mT (in between the third and fourth panels of Fig. S2). In the region close to the center of the sample where switching normally occurs (labelled "Mixed"), we still see oscillations in the y-component, but mainly noise in the x and z components. This is to be expected if the number of VC reversals varies stochastically from one magnetic field pulse to the next. Regardless of the sense of gyration (CW or CCW) the trajectory component $y(t)$ is unchanged, and therefore still shows clear oscillations. However, the x-component will be reversed if the sense of gyration is reversed. For example, at the moment the pulse is turned off and the VC is at a position $y < 0$, we would expect motion in the positive x-direction for CW motion and motion in the negative x-direction for CCW direction. And of course, we expect the z-component to be reversed for opposite core polarity. Yet both CW and CCW motion would result in motion in the positive y-direction. Therefore, if the number of switches varies from one pulse to the next, we expect the x- and z- components to average to zero, but the y-component to still display clear oscillations. Interestingly, the oscillations are still visible in all components in the dynamics during the pulse. This is expected if no stochastic switching occurs in this region, though it also requires that the VC begins with some preferred polarity. We believe that this region of non-deterministic switching occurs as the VC is in between regions of no switching and deterministic switching, and thermal or electronic noise is sufficient to cause some of the VC reversals to either occur or not from one repetition to the next.

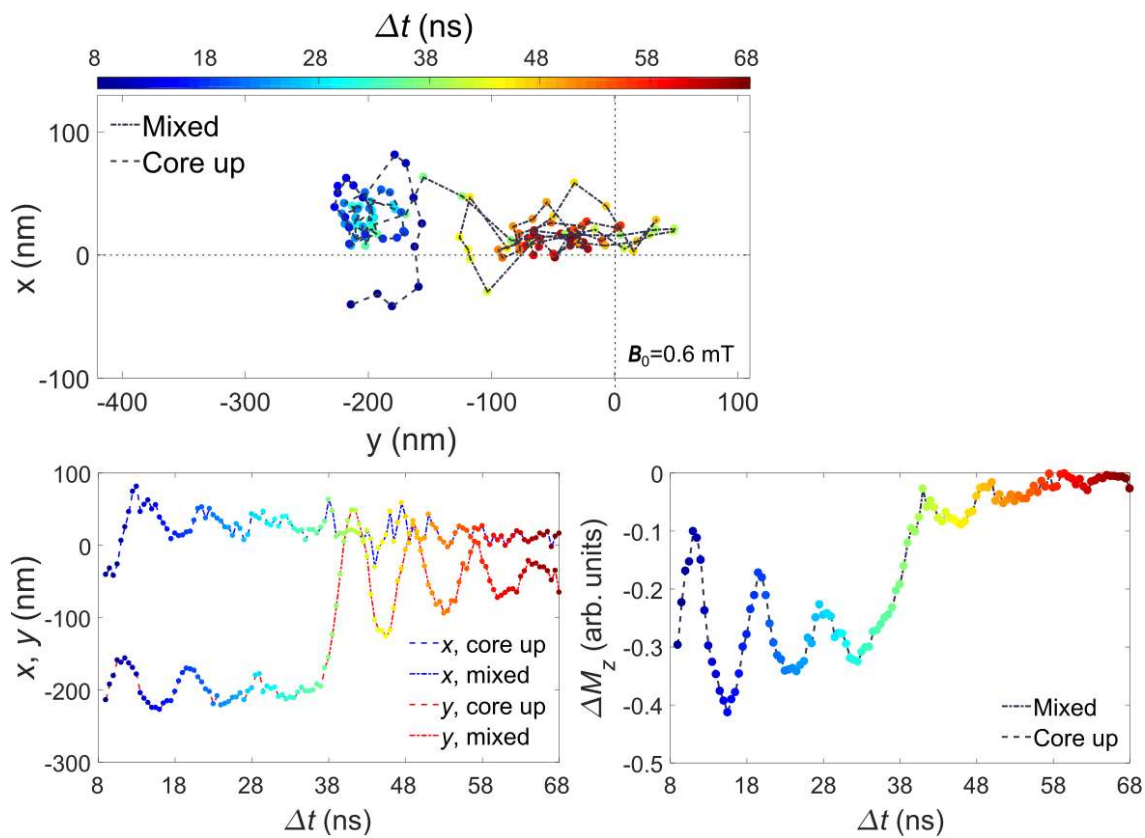

**Fig. S3.** Region of stochastic switching (sample 1). The top panel demonstrates the 2D gyrations of the VC at $B_0 = 0.6$ mT. The x and y components are shown separately vs. time on bottom left. Bottom right illustrates the change in the z-component. To excite these gyrations a 3.2 mT amplitude, and 35 ns magnetic pulse is used. Color bar $\Delta t$ indicates the time relative to the beginning of the pulse.

In order to check the polarity flip at smaller field pulses, we used a 1.6 mT pulse on the first sample to check the gyrations close to the center. Figure S4 shows the 3D TR-MOKE results. Here, we used no offset field, and the pulse duration is 35 ns. In this figure, there are at least three switching spots. Most clearly, switching can be seen near $(x, y) = (0, \pm 50)$ nm when the VC starts to relax back to the center. It is possible that switching may be possible at still lower pulse fields, though the signal becomes increasingly difficult to discern above the noise.

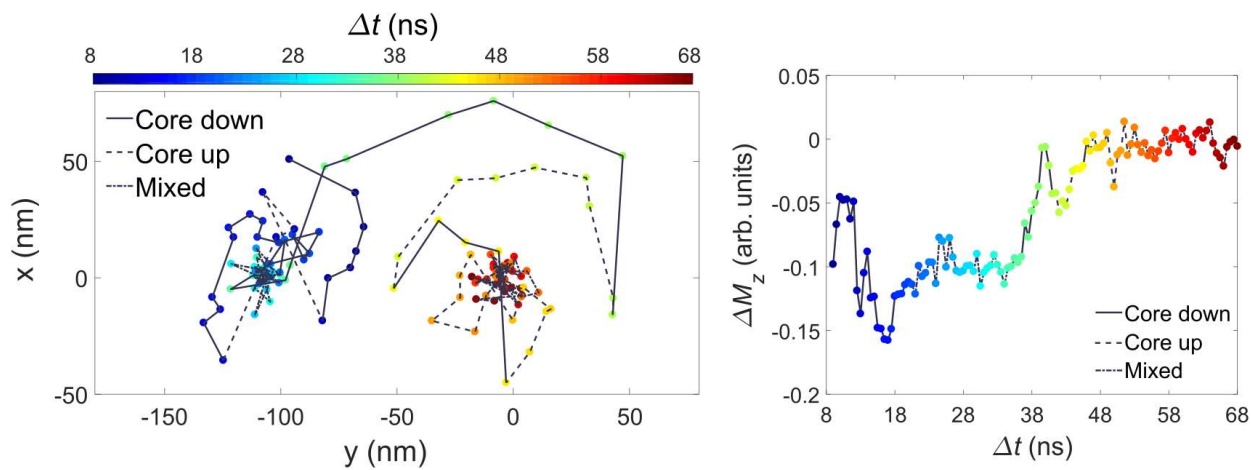

**Fig. S4.** Flipping the core polarity at lower magnetic fields. Left panel shows the 2D image for the VC gyrations and the right panel shows the change in the z-component. In this case a 1.6 mT amplitude, 35 ns magnetic pulse is used. Color bar $\Delta t$ indicates the time relative to the beginning of the pulse.

To demonstrate single-shot switching, we started from a new sample with no observed switching (Fig. S5).

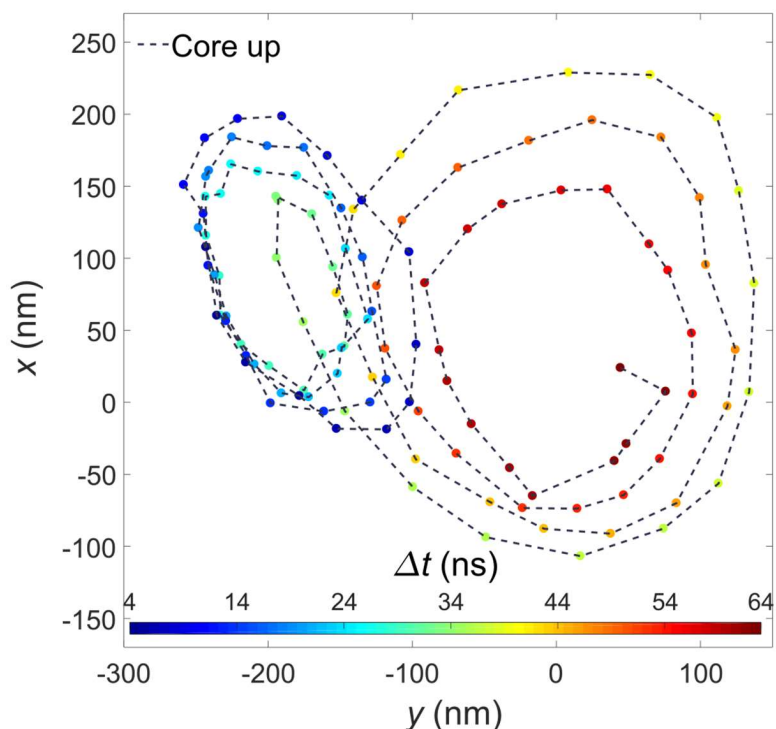

**Fig. S5.** Normal gyrations on disk 2 before the prolonged pulsed laser exposure. This scan is an average of two scans. The color bar $\Delta t$ shows the time relative to the beginning of the 35 $ns$ pulse.

We then exposed the sample to the pulsed laser, while repeatedly measuring the VC trajectory until we observed switching events begin to appear. The first switching spot observed is shown in Fig. S6. The second switching spot observed is shown in the main text in Fig. 3(**A**). We use the second switching spot for the single shot switching, because we were unable to find a clear region with no switching sufficiently close to that the first spot as to be within the range of the pulse generator, where we could test the subsequent VC polarity.

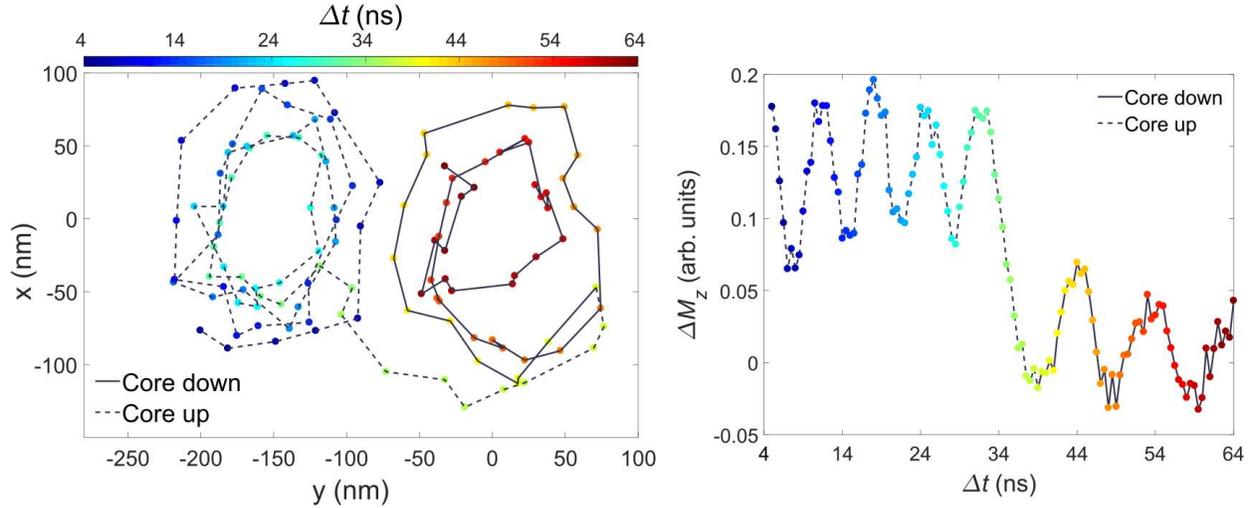

**Fig. S6.** First switching observed in sample two. The left panel demonstrates the 2D image of gyrations, and the right panel illustrates the z-component oscillation. In this case a 2.1 mT amplitude, 35 ns magnetic pulse is used. Color bar $\Delta t$ indicates the time laps after the pulse is turned on.

To implement the single shot switching, we find a magnetic field amplitude and offset such that a 5-ns-duration pulse sends the VC to the switching site. Following a single-shot pulse, we translate the VC slowly (over several microseconds) to a new equilibrium position where we still observe dynamics with no VC reversals. A short TR-MOKE scan about this new equilibrium then reveals the VC polarity, as shown in Fig. S7. The scans in Fig. S7 are obtained with significantly more averaging, so as to reduce the noise for demonstration purposes; in practice, faster scans suffice to determine the gyration direction.

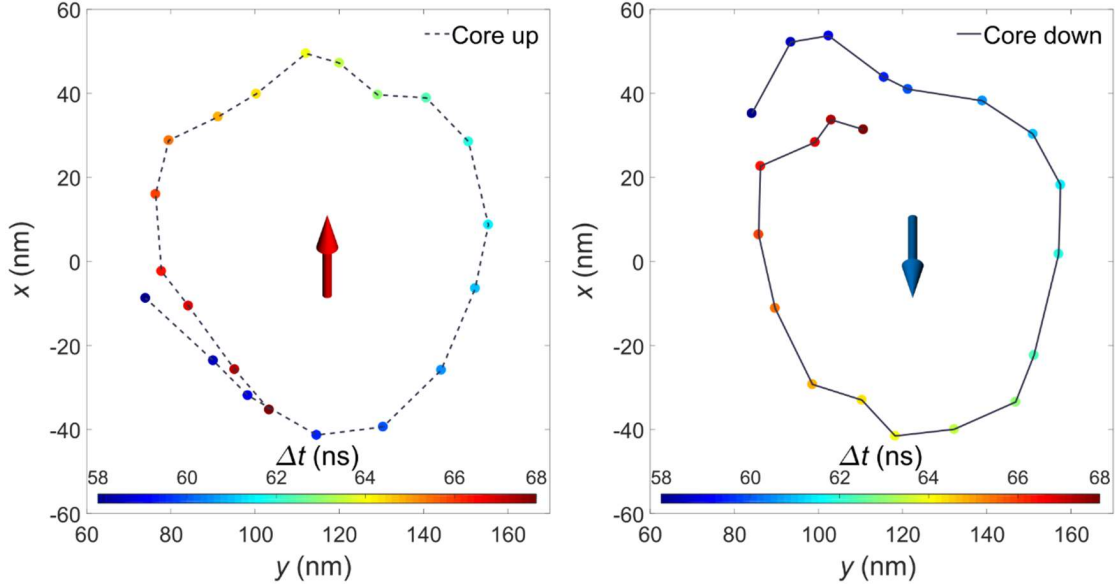

**Fig. S7.** Examples of VC dynamics used to measure core polarities up and down respectively after single-shot switch. The color bars $\Delta t$ show the time relative to the beginning of the 35 $ns$ pulse.

**Simulation methods:**

To reveal the physics underlying the core polarity switching, we performed micromagnetic simulations using an in-house code (*19*). We used a disk of with a diameter of 1,000 nm and a thickness of 40 nm on a regular grid of $2 \times 2 \times 5$ nm$^3$. We generally used this disk diameter rather than the 2,200 nm of the experimental systems in order to reduce simulation times; in any case, the physical mechanism of core polarity reversal does not depend on the disk diameter so long as it is much larger than both the magnetic exchange length and the disk thickness. The code integrates the Landau-Lifshitz-Gilbert equations $\frac{d\boldsymbol{m}}{dt} = -\frac{\gamma}{1+\alpha^2}\boldsymbol{m}\times\boldsymbol{H}_{eff} - \frac{\gamma\alpha}{1+\alpha^2}[\boldsymbol{m}\times(\boldsymbol{m}\times\boldsymbol{H}_{eff})]$, with $\boldsymbol{m}$ the magnetization director, and the effective field $\boldsymbol{H}_{eff} = -\frac{\delta E}{\delta M}$, with $E$ the total energy and $M$ the magnetization density. The total energy includes exchange energy, magnetostatic energy, anisotropy energy (here negligible), and Zeeman energy. The magnetostatic field is calculated using fast Fourier transform, and time-integration is done using an implicit adjustable-time Bulirsch-Stoer integrator.

We first performed quasi-static simulations of disks in an in-plane magnetic field of 2.5 – 5.0 mT and calculated the displacement of the vortex core using a Gilbert damping of $\alpha$=0.2. For a disk of diameter 2,000 nm and magnetic parameters *A*=13 pJ/m and a magnetization density of M=8.00×10$^5$ A/m we obtained a core displacement of about 52 nm/mT, and for a diameter of 1,000 nm we obtained about 14 nm/mT, which are in reasonable range of published values (*20*); for a 2,200 nm diameter disk 40 nm thick with M=7.80×10$^5$ (7.00×10$^5$) A/m and a $2 \times 2 \times 10$ nm$^3$ mesh we obtained a displacement of 64 (72) nm/mT. The experimentally estimated displacement susceptibilities are 66 – 76 nm/mT, and we therefore generally reduced the magnetization density to 7.00×10$^5$ A/m in our simulations. The exchange coupling is less important for the core displacement as it is primarily driven by magnetostatic interactions, and in the simulations below we used values of *A*=10 pJ/m and *A*=13 pJ/m.

Next, we wanted to verify that defect areas in which the exchange coupling in which the exchange coupling has been reduced, e.g. by diffusion of oxygen or other defects from the Py-Au interface, can give rise to the observed hysteresis and core polarity reversal.

We first created a "defect region" of radius 15 nm centered 115 nm from the disk center along the y-axis. In this region, we reduced the exchange coupling $A$ from $A=13$ pJ/m in the disk to values ranging from 1.3 to 10 pJ/m, and for each value of the exchange coupling simulated the quasi-static trajectory of the vortex core in applied fields from 0 mT to 15 mT and then back to 0 mT along the x-axis. Traces of the vortex core are shown in the main text in Fig. 4(A). At $A=10$ pJ/m, there is a small onset of hysteresis visible, and at $A=2.5$ pJ/m the hysteresis is quite large. The core is deformed and deflected by the defect region towards negative x with increasing field, and towards positive x with decreasing field (not shown in the figure). At the lowest value of $A$, 1.3 pJ/m, the lateral deflection is about 27 nm, while for $A=10$ pJ/m, the deflection is only about 5 nm with no clear deformation of the core.

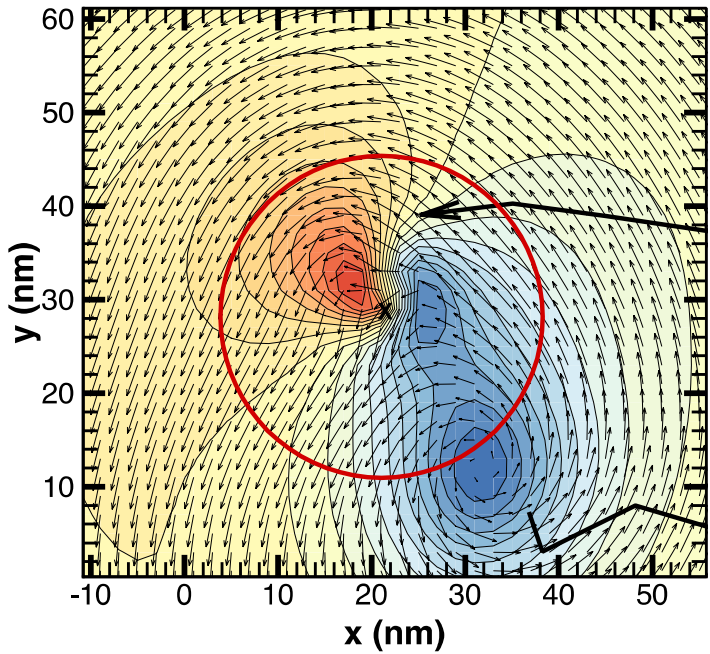

**Fig. S8** Enlarged detail of vortex core and wake just before the initial vortex core (in red) is annihilated.

In order to simulate dynamical core reversal, we created a defect region of radius 150 nm centered 115 nm from the disk center along the y axis. In this region, we reduced the magnetization density to $5.00\times10^5$ A/m and set the exchange coupling to between 1.3 pJ/m to 10 pJ/m with $A=10$ pJ/m in the rest of the disk. We then initialized the system with a vortex with a core located about 150 nm below the disk center and time-integrated the magnetization evolution in time steps of 0.25 ps with a Gilbert damping of $\alpha=0.02$. For exchange couplings larger than about 3.5 pJ/m, the vortex core does not reverse polarity. However, for $A=3.5$ pJ/m and smaller, the core reverses polarity as it enters the defect region. Based on the theory of core critical velocity (3, 4), the critical velocity is proportional to the square root of the exchange coupling, $v_c \propto \sqrt{A}$. In our simulations, the core velocity is about 200 m/s. We can then estimate the critical exchange coupling $A_c$ below which the core polarity flips to be about 4.8 pJ/m, which is in rather good agreement with the numerical simulations. We note that when the exchange coupling becomes too small, the core disintegrates when it enters the defect region; this was the case for $A$=1.3 pJ/m in our simulations. We also performed some checks that the reduction of the magnetization density did not appreciably change our conclusions by running simulations with $M$=7.00×10$^5$ A/m in the defect region. Finally, we also checked that the initial core polarity did not alter our conclusions by running simulations reversing the initial polarity.

For a closer look at the reversal process itself, Fig. S8 shows an enlarged detail of the initial VC (red) and enhanced wake (blue) just before annihilation of the initial VC. The wake has deepened, and split into two pieces: one annihilates with the VC, and the other splits off and becomes a new VC. This figure is at 1.4 ns after Fig. 1B in the main manuscript. The color scale is the same as in Figs. 4B-F, and the vector arrows indicate the in-plane magnetization.

## Movies S1-S2

**Movie S1:** Movie of the vortex state in quasi-equilibrium as an applied field is swept up and down, translating the VC near a small region with reduced $A = 2.5$ pJ/m, relative to $A = 13$ pJ/m elsewhere. The VC becomes trapped up against the defect region, resulting in hysteretic VC position vs. applied field.

**Movie S2:** Movie of VC reversal process upon encountering a defect region. VC is initially undergoing CCW gyrotropic motion in the normal region, with $A = 10$ pJ/m. The VC then enters a circular region (shown by the dashed circle), with reduced $A = 3.5$ pJ/m. At the spot marked by an X in a red circle, the original VC is annihilated in a burst of spin waves, and the reversed VC begins following a CW path.